\documentclass[10pt]{iopart}
\usepackage{iopams}
\usepackage{setstack}
\usepackage[margin=2cm]{geometry}
\usepackage{float}
		\usepackage{graphicx,color}
		\graphicspath{{2017-12-05_figures/}}
		\usepackage{enumerate}
		\usepackage{epstopdf}
		\usepackage{textcomp}
		\usepackage{gensymb}
		
		\newcommand{\moy}[1]{\left\langle #1 \right\rangle}

		\newcommand{\ex}[1]{\mathrm{e}^{#1}}

		\newcommand{\dd}[0]{\mathrm{d}}

		\newcommand{\RR}[0]{\boldsymbol{R}}
		\newcommand{\xx}[0]{\boldsymbol{x}}

		\newcommand{\pp}[0]{\boldsymbol{p}}
		
		\newcommand{\kB}[0]{k_{\mathrm{B}}}
		
		\newcommand{\nn}[0]{\hat{\boldsymbol{n}}}
		
		\newcommand{\ep}[0]{\epsilon}

		\newcommand{\rotop}[0]{\boldsymbol{\mathcal{R}}}
		\newcommand{\uu}[0]{ \hat{\boldsymbol{u}}}
		
\begin{document}

\title{Fluctuation-induced hydrodynamic coupling in an asymmetric, anisotropic dumbbell}
\author{Tunrayo Adeleke-Larodo$^1$, Pierre Illien$^{1,2,3}$\footnote{Present address:
EC2M, UMR Gulliver, ESPCI Paris, 10 rue Vauquelin, 75005 Paris, France} and Ramin Golestanian$^{1,3}$}
\address{$^1$Rudolf Peierls Centre for Theoretical Physics, University of Oxford, Oxford OX1 3NP, UK}
\address{$^2$Department of Chemistry, The Pennsylvania State University, University Park, PA 16802, USA}
\address{$^3$Max Planck Institute for the Physics of Complex Systems, N${\ddot{\rm o}}$thnitzer Str. 38, D-01187 Dresden, Germany}

\ead{ramin.golestanian@physics.ox.ac.uk}
\vspace{10pt}
\begin{abstract}
We recently introduced a model of an asymmetric dumbbell made of two hydrodynamically coupled subunits as a minimal model for a macromolecular complex, in order to explain the observation of enhanced diffusion of catalytically active enzymes. It was shown that internal fluctuations lead to negative contributions to the overall diffusion coefficient and that the fluctuation-induced contributions are controlled by the strength of the interactions between the subunits and their asymmetry. We develop the theory further by studying the effect of anisotropy of the constituents on the diffusion properties of a modular structure. We derive an analytic form for the diffusion coefficient of an asymmetric, anisotropic dumbbell and show systematically its dependence on the internal and external symmetry. We give expressions for the associated polarisation fields, and comment on their consequences for the alignment mechanism of the dumbbell. The present work opens the way to more detailed descriptions of the effect of hydrodynamic interactions on the diffusion and transport properties of biomolecules with complex structures.
\end{abstract}
\maketitle
\section{Introduction}
Understanding the dynamics of biomolecules is crucial if we are to fully appreciate and potentially harness their functionalities in the search for biocompatible micro- and nano-machines. Enzymes, in particular, have the ability to perform very specific functions under conditions dominated by thermal fluctuations and viscous hydrodynamics \cite{alberts_2014}, and yet there is relatively little known about the physical characteristics of this remarkable class of biomolecules. Recently, the question of the effect of catalytic activity of enzymes on their diffusion properties has been addressed: Experiments with dilute solutions of enzyme molecules which catalyse exothermic reactions and typically have high catalytic rates have revealed that their diffusion is substantially enhanced in a substrate-dependent manner when they are catalytically active \cite{muddana_2010,sengupta_2014, sengupta_2013,riedel_2014}.\par
The observations were followed by theoretical investigations into the underlying mechanism of the phenomenon of enhanced diffusion. Initial explanations relied on the exothermicity and fast rate of catalysis of the chemical cycle of an enzyme \cite{switala_2002,golestanian_2015,riedel_2014}. The effect of hydrodynamic coupling of enzyme molecules to their environment was also studied \cite{golestanian_2015,mikhailov_2015,golestanian-ajdari_2008,cressman_2008,bai_2015}. These theories were based on the key idea that the non-equilibrium catalytic cycle is the main driving force behind the observed phenomenon. However, as it later turned out, the observations appear to be independent of the thermodynamic properties and the non-equilibrium activity of an enzyme; recent experiments have demonstrated similar levels of enhancement of the diffusion coefficient of the endothermic and slow enzyme aldolase \cite{illien_exothermicity_2017}. In \cite{illien_diffusion_2017}, we proposed a new theoretical approach for understanding the phenomenon, capable of replicating all experimental observations. We introduced the classical dumbbell model as a minimal model to study the effect of hydrodynamic interactions and catalytic activity of enzyme molecules, and the consequence on their diffusivity.\par
The effect of hydrodynamic interactions on the dynamics of macromolecular suspensions is a long-standing problem of polymer physics \cite{doi_theory_1986}, whose study requires approaches from statistical physics, low Reynolds number hydrodynamics and rheology. Models of flexible chains or dumbbells have been extensively studied and successfully used to model the dynamics of real macromolecules. Equilibrium-like averaging procedures, pioneered by Zimm \cite{zimm_dynamics_1956} and subsequently refined \cite{ottinger_rouse_1989,ottinger_dumbbell_1989}, have been proposed to account for hydrodynamic interactions. The so-called `pre-averaging' strategies are known to be very accurate \cite{doi_theory_1986}. Numerical strategies allowing efficient sampling of the configurations of such model polymers were employed to study, for instance, the behaviour of chains and dumbbells under shear \cite{doi_theory_1986}, their cyclisation dynamics \cite{wilemski_1974,levernier_2015} or their diffusion properties. In Ref. \cite{illien_diffusion_2017}, we investigated effects that were overlooked so far, such as the effect of {\em internal asymmetries}, inherent in real macromolecules and in particular in enzymes which bear an active site and therefore a built-in asymmetry, the effect of {\em orientation fluctuations} of the subunits that constitute the dumbbell that couple to the compressional degrees of freedom, and the effect of {\em changes between the different conformational landscapes} explored by the dumbbell during its catalytic cycle.\par
In Ref. \cite{illien_diffusion_2017}, the hydrodynamic tensors were approximated by the isotropic part which amounts to a pre-averaging of the orientation dependence, a common method in studies of the hydrodynamic properties of macromolecules in solution. However, orientation-pre-averaging compromises knowledge of the effect of local anisotropy. This then raises questions on the effect of the inclusion of anisotropy on the dynamics of the system and the problem of a consistent treatment of the orientation dependence. Here we present resolutions by considering a more general case of the model introduced in Ref. \cite{illien_diffusion_2017} through the inclusion of hydrodynamic interactions due to anisotropy, with the aim of completing our description. In so doing, we provide a comprehensive analytical theory for the dynamics of modular structures, and we open the way to a better description of the effect of hydrodynamic interactions on the diffusion and transport properties of biomolecules with complex structures.
\section{The model}
		In order to study the effects of fluctuation-induced hydrodynamic coupling, we consider the simplest system, endowed with a minimal number of degrees of freedom, exhibiting this type of phenomenon. We consider a pair of rigid, unequal Brownian particles which are of arbitrary shape, suspended in an unbounded fluid. The particles are coupled through hydrodynamic interactions and through an interaction potential $U$. We assume Stokes flow, so that the forces and torques exerted by the subunits on the fluid are linearly related to the instantaneous linear and angular velocities through hydrodynamic interactions \cite{happel1973low,kim_microhydrodynamics:_2005}. In addition, their positions $\xx^\alpha$ and orientations $\uu^\alpha$ undergo thermal fluctuations around the equilibrium configuration. As such, the system describes low Reynolds number flow around an asymmetric dumbbell which accesses different compressional and orientation modes in a fluctuation-dependent way. In the experiments reporting enhanced diffusion, the concentration of enzymes were very low (10 nM in \cite{illien_exothermicity_2017}), corresponding to a small volume fraction. This justifies our expectation of describing the phenomenon with a single enzyme.\par
		The interaction potential $U$ is defined through the enabled modes of fluctuation: $\uu^1$ and $\uu^2$ are taken to be coplanar and only fluctuations within the plane are considered (this simplification is justified on the premise that fluctuations out of the plane are expected to be substantially weaker), specifically, our focus is on fluctuations in the orientations $\uu^\alpha$, fluctuations about the other two body axes of each subunit do not feature explicitly in our calculations, though their inclusion is through a simple extension of our analysis, but no new phenomena would be introduced. Therefore, the interaction potential may be expressed as a Legendre expansion in the angles between the orientation vectors $\uu^1$, $\uu^2$ and $\nn$ as
		\begin{equation}
		U = V_0(x)+\sum_{\alpha=1,2}V_{\alpha}(x) \,\nn \cdot \uu^\alpha + V_{12}(x) \,\uu^1 \cdot \uu^2		\label{potential} 
		\end{equation}
		at first order. The functions $V_0(x)$, $V_{\alpha}(x)$ and $V_{12}(x)$ are generic and distinct. $V_0(x)$ contributes only to the extension of the dumbbell, while $V_{\alpha}(x)$ and $V_{12}(x)$ in addition quantify the strength of the constraints on orientation fluctuations.\par
				\begin{figure}
\centering
\includegraphics[width=0.3\textwidth]{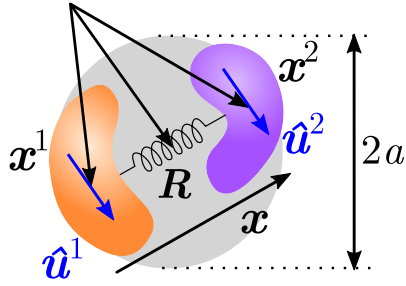}
\caption{The generalised dumbbell, which has typical size $a$ and consists of two non-identical subunits at positions $\xx^\alpha$ and orientations $\uu^\alpha$ which are hydrodynamically coupled and also interact via some potential $U$. The centre of mass of the dumbbell is denoted by $\mathbf{R}$ and the elongation (i.e., separation of the subunits) by $\xx$. The subunits fluctuate about the equilibrium configuration due to thermal fluctuations.}
\label{fig:dumbell}
\end{figure}
			\section{Smoluchowski equation}
		We begin our analysis with the Smoluchowski equation for a pair of interacting Brownian particles. The probability $P(\xx^1 ,\xx^2 ,\uu^1, \uu^2; t)$ of finding subunit $\alpha$ at position $\xx^\alpha$ and with orientation $\uu^\alpha$ at time $t$ has the following evolution equation
		\begin{eqnarray}
		\partial_t P =& \sum_{\alpha,\beta=1,2}& \Big\{\nabla_\alpha \cdot  \mathbf{M}_{\mbox{\scriptsize TT}}^{\alpha\beta} \cdot \left[(\nabla_\beta U)P + \kB T \nabla_\beta P\right] + \nabla_\alpha \cdot \mathbf{M}_{\mbox{\scriptsize TR}}^{\alpha\beta} \cdot\left[(\rotop^\beta  U) P + \kB T \rotop^\beta P \right] \nonumber\\
		 &+&  \rotop^\alpha \cdot \mathbf{M}_{\mbox{\scriptsize RT}}^{\alpha \beta}  \cdot \left[(\nabla_\beta U) P + \kB T \nabla_\beta P\right] + \rotop^\alpha \cdot \mathbf{M}_{\mbox{\scriptsize RR}}^{\alpha \beta}  \cdot \left[(\rotop^\beta U) P + \kB T \rotop^\beta P\right] \Big\},
		\label{smoluchowski}
		\end{eqnarray}
		where $\mathbf{M}_{\mbox{\scriptsize AB}}^{\alpha\beta}$ are elements of a mobility matrix which couples the interactions between the translational (T) and rotational (R) modes of the subunits. 
For $\alpha \neq \beta$, $\mathbf{M}_{\mbox{\scriptsize AB}}^{\alpha\beta}$ correspond to the hydrodynamic interactions between the subunits, for $\alpha=\beta$ it represents self-mobility in the presence of the other subunit. The mobility matrix is symmetric and positive definite if the interacting particles are identical \cite{doi_theory_1986}.\par
The interaction potential is $U(\xx^1 ,\xx^2 ,\uu^1, \uu^2)$ as before and $\rotop^\alpha = \uu_\alpha \times \partial _{\uu_\alpha} $ is the rotational gradient operator \cite{doi_theory_1986}. Equation (\ref{smoluchowski}) is known in polymer dynamics as the equation of motion for dilute polymer solutions as presented in the Zimm model \cite{doi_theory_1986}. However, here the constituents are unequal and non-axisymmetric.\par
For convenience we separate the dynamics into pure translation, where the translational modes of the subunits are coupled, and rotation, which incorporates coupling of rotational modes and also cross coupling of rotation and translation. Equation (\ref{smoluchowski}) can then be written as
		\begin{equation}
		\label{operatorFP_1}
		\partial_t P = \mathcal{L}_{\mbox{\scriptsize T}} P + \mathcal{L}_{\mbox{\scriptsize R}} P,
		\end{equation}
		where
		\begin{eqnarray}
		\label{operatorFP_2}
		 \mathcal{L}_{\mbox{\scriptsize T}} P = &\sum_{\alpha,\beta=1,2} &  \nabla_\alpha \cdot \mathbf{M}_{\mbox{\scriptsize TT}}^{\alpha\beta} \cdot  \left[  (\nabla_\beta  U) P + \kB T \nabla_\beta P \right], \nonumber\\
		 \mathcal{L}_{\mbox{\scriptsize R}} P =& \sum_{\alpha,\beta=1,2} & \left(\nabla_\alpha \cdot \mathbf{M}_{\mbox{\scriptsize TR}}^{\alpha\beta} + \rotop^\alpha \cdot \mathbf{M}_{\mbox{\scriptsize RR}}^{\alpha \beta}\right)  \cdot  \left[  (\rotop^\beta  U) P + \kB T \rotop^\beta P \right] + \rotop^\alpha \cdot \mathbf{M}_{\mbox{\scriptsize RT}}^{\alpha \beta}  \cdot [(\nabla_\beta U) P + \kB T \nabla_\beta P].
		\end{eqnarray}
		In the centre-of-mass $\RR = (\xx^1+\xx^2)/2$ and separation $\xx = \xx^2-\xx^1=x\nn$ coordinates of the dumbbell, (\ref{operatorFP_2}) becomes
		\begin{equation}
		\label{LTP}
		\mathcal{L}_{\mbox{\scriptsize T}} P = \left(\frac{1}{4} \nabla_{\RR} \cdot \mathbf{M} +\frac{1}{2}\nabla_{\xx} \cdot \boldsymbol{\Gamma} \right) \cdot \kB T\nabla_{\RR} P+ \left(\frac{1}{2}\nabla_{\RR} \cdot \boldsymbol{\Gamma} + \nabla_{\xx} \cdot \mathbf{W}\right) \cdot \left[(\nabla_{\xx}U)P + \kB T \nabla_{\xx} P\right],
		\end{equation}
		and
		\begin{eqnarray}
		\mathcal{L}_{\mbox{\scriptsize R}} P = & \sum_{\alpha,\beta=1,2}& \Bigg\{\left(\frac{1}{2} \nabla_{\RR} \cdot\boldsymbol{\Phi}^{(\alpha)} + (-1)^\alpha\nabla_{\xx}\cdot\boldsymbol{\Lambda}^{(\alpha)}\right) \cdot \left[\kB T \rotop^\alpha P + (\rotop^\alpha U)P\right]+\frac{1}{2}\rotop^\alpha\cdot\boldsymbol{\Phi}^{(\alpha)} \cdot \kB T\nabla_{\RR} P\nonumber\\
		&+&(-1)^\alpha\rotop^\alpha\cdot \boldsymbol{\Lambda}^{(\alpha)} \cdot[(\nabla_{\xx}U)P + \kB T \nabla_{\xx} P]+\rotop^\alpha \cdot \mathbf{X}^{\alpha\beta}\cdot[\kB T\rotop^\beta P + (\rotop^\beta U)P]\Bigg\}.
		\label{LRP}
		\end{eqnarray}
		 Using the symmetry property $\mathbf{M}^{\alpha\beta}_{\mbox{\scriptsize AB}}=\mathbf{M}^{\beta\alpha}_{\mbox{\scriptsize BA}}$ (which follows from the Lorentz reciprocal theorem \cite{happel1973low,kim_microhydrodynamics:_2005}), the new Smoluchowski equation is written in terms of new hydrodynamic tensors which are linear combinations of the old ones, so that $\mathbf{M}$, $\mathbf{W}$ and $\boldsymbol{\Gamma}$ are composed of the translation tensors, $\boldsymbol{\Psi}^{(\alpha)}$ and $\mathbf{X}^{\alpha\beta}$ the rotation tensors, and $\boldsymbol{\Lambda}^{(\alpha)}$ and $\boldsymbol{\Phi}^{(\alpha)}$ correspond to translation-rotation coupling. We continue our analysis by exploiting the difference in scale between the centre-of-mass and separation coordinates.
		 		\section{Averaging procedure}
		The diffusion coefficient of the centre-of-mass of the dumbbell is given by an average over the configurations, hence $D_{\mbox{\scriptsize eff}}= \lim_{t\to\infty}\frac{1}{6} \frac{\dd}{\dd t} \int_{\RR} \int_{\xx} \int_{\uu^1} \int_{\uu^2} {\RR^2 P}$. The calculation of $D_{\mbox{\scriptsize eff}}$ in this way, using the Smoluchowski equation (equations (\ref{LTP}) and (\ref{LRP})) involves higher-order correlation functions which may combine the external ($\RR$) and internal ($\xx, \uu^1$ and $\uu^2$) degrees of freedom. Furthermore, expressions  for these correlation functions calculated using (\ref{LTP}) and (\ref{LRP}) yield yet higher-order moments. This hierarchy is closed through the definition of an averaging procedure which eventually leads to a closed expression after a suitable truncation approximation.\par
		A well-defined averaging procedure is motivated by the identification of a separation of the time-scales in the dynamics of the dumbbell: There are three time-scales, each describing the relaxation time of a degree of freedom. The slowest of the three is the relaxation time of the centre-of-mass coordinate $\RR$. Given a potential that can be Taylor expanded around the minimum, the quadratic term gives the effective spring constant of the potential ($k$) and the time-scale for $x$ to return to its equilibrium value, $\tau_s=\xi/k$, where $\xi$ is the friction coefficient of the enzyme. The rotational diffusion time of the enzyme $\tau_r$, which determines the rate of loss of memory of orientation through $\moy{\nn(t)\cdot\nn(0)}=\exp(-|t|/\tau_r)$, is of the order $\xi a^2/\kB T$. The ratio of the two times $\tau_s/\tau_r$ goes as $\kB T/ka^2\sim\delta x/a$, the relative deformation of the enzyme due to thermal fluctuations, and is therefore bounded by unity. With this consideration, we can average over the separation of the subunits assuming $\nn$, $\uu^1$ and $\uu^2$ to be fixed. The average is defined as $\moy{\cdot}=\frac{1}{\mathcal{P}}\int \dd x\,x^2 \cdot P$, where $\mathcal{P}=\int \dd x\,x^2 \cdot P$. \par
		The average over the separation $x$ of the Smoluchowski equation requires expressions for the gradients of the interaction potential, $\nabla_{\xx} U$ and $\rotop^\alpha U$. Applying $\nabla_{\xx} = \nn \partial_x - \frac{1}{x}\nn\times \rotop$ and $\rotop^\alpha$ to (\ref{potential}) gives
		\begin{equation}
		\label{grad_x_U}
		\nabla_{\xx} U = \nn U' + (\mathbf{1}-\nn\nn)\cdot \bigg(\frac{V_1(x)}{x}\uu^1+\frac{V_2(x)}{x}\uu^2\bigg),
		\end{equation}
		and 
		\begin{equation}
		\rotop^\alpha U = V_{12}(x)(\uu^\alpha\times\uu^\beta) + V_{\alpha}(x)(\uu^\alpha\times\nn).
		\label{rotop_U}
		\end{equation}
	 Furthermore, for a hydrodynamic tensor $\mathbf{A}$, $\nabla_{\xx}\cdot\mathbf{A} = \nabla_{\xx}\cdot\left(\nn\nn\cdot\mathbf{A} + (\mathbf{1}-\nn\nn)\cdot\mathbf{A}\right)$, has just one contribution $-(\nn\times \rotop)\cdot(\mathbf{1}-\nn\nn)\cdot\frac{\mathbf{A}}{x}$ after averaging over the separation. The full expression for the evolution of the separation-averaged distribution $\mathcal{P}(\RR,\nn,\uu^1,\uu^2;t)$ is given by the sum of the following
	 \begin{eqnarray}
		\mathcal{L}_{\mbox{\scriptsize T}}\mathcal{P} & = & \left[\frac{1}{4} \nabla_{\RR} \cdot \moy{\mathbf{M}} -\frac{1}{2} (\nn\times\rotop)\cdot(\mathbf{1}-\nn\nn)\cdot \moy{\frac{\boldsymbol{\Gamma}}{x}}\right]\cdot \kB T\nabla_{\RR} \mathcal{P}\nonumber \\
		& & - \frac{\kB T}{2} \partial_{R_i} \epsilon_{jkl} \Bigg[\mathcal{R}_l\Bigg( \moy{\frac{\Gamma_{ij}}{x}}\hat{n}_k\mathcal{P}\Bigg) - \moy{\mathcal{R}_l \bigg(\frac{\Gamma_{ij}}{x}\hat{n}_k\bigg)}\mathcal{P} \Bigg]\nonumber\\
		& & + \kB T (\nn\times\rotop)_i(\mathbf{1}-\nn\nn)_{ij}\ep_{klm}\Bigg[\mathcal{R}_m\Bigg(\moy{\frac{W_{jk}}{x^2}} \hat{n}_l\mathcal{P}\Bigg)- \moy{\mathcal{R}_m \Bigg(\frac{W_{jk}}{x^2}\hat{n}_l\Bigg)}\mathcal{P}\Bigg] \nonumber \\
		& &  + \left[\frac{1}{2} \nabla_{\RR} \cdot \sum_{\alpha=1,2}\bigg<\frac{\boldsymbol{\Gamma} \, V_{\alpha}(x)}{x}\bigg> - (\nn\times\rotop) \cdot (\mathbf{1}-\nn\nn)\cdot \sum_{\alpha=1,2}\bigg<\frac{\mathbf{W} \, V_\alpha(x)}{x^2}\bigg>\right] \cdot (\mathbf{1}-\nn\nn) \cdot \uu^\alpha \mathcal{P},
		\label{LTPav}
		\end{eqnarray}
		and
		\begin{eqnarray}
		\mathcal{L}_{\mbox{\scriptsize R}}\mathcal{P} & = & \sum_{\alpha,\beta=1,2}\Bigg\{\frac{\kB T}{2}\partial_{\mathcal{R}_i}\left[\mathcal{R}^\alpha_{j} (\langle \Phi^{(\alpha)}_{ij} \rangle \mathcal{P}) - \moy{\mathcal{R}^\alpha_{j} \Phi^{(\alpha)}_{ij}} \mathcal{P}\right] + \kB T\mathcal{R}^\alpha_{i}\left[\mathcal{R}^\beta_{j} \left(\langle X^{\alpha\beta}_{\mbox{\scriptsize RR}}\;_{ij} \rangle \mathcal{P}\right) - \langle \mathcal{R}^\beta_{j} X^{\alpha\beta}_{\mbox{\scriptsize RR}}\;_{ij} \rangle \mathcal{P}\right]\nonumber \\
		 &&+ \frac{1}{2}\nabla_{\RR} \cdot\left[\langle\boldsymbol{\Phi}^{(\alpha)} V_{12} \rangle \cdot (\uu^\alpha \times \uu^\beta)\mathcal{P} + \langle\boldsymbol{\Phi}^{(\alpha)} V_{\alpha} \rangle \cdot (\uu^\alpha \times \nn)\mathcal{P}\right]\nonumber \\
		&&- \boldsymbol{\mathcal{R}}^\alpha \cdot\left[\langle\mathbf{X}_{\mbox{\scriptsize RR}}^{\alpha\beta} V_{12} \rangle\cdot (\uu^\alpha \times \uu^\beta)\mathcal{P} + \langle\mathbf{X}_{\mbox{\scriptsize RR}}^{\alpha\beta} V_{\beta} \rangle\cdot (\uu^\beta \times \nn)\mathcal{P}\right]\nonumber\\
		&& + \boldsymbol{\mathcal{R}}^\alpha \cdot\left[\langle\boldsymbol{\Phi}^{(\alpha)} V_{12} \rangle \cdot (\uu^\alpha \times \uu^\beta)\mathcal{P} + \langle\boldsymbol{\Phi}^{(\alpha)} V_{\alpha} \rangle \cdot (\uu^\alpha \times \nn)\mathcal{P}\right]\nonumber\\
		&& -\kB T(-1)^\alpha\mathcal{R}^\alpha_{i} \ep_{jkl}\left[\mathcal{R}_l \left(\moy{\frac{\boldsymbol{\Lambda}^{(\alpha)}_{ij}}{x}} \hat{n}_k \mathcal{P}\right) - \moy{\mathcal{R}_l\left(\frac{\boldsymbol{\Lambda}^{(\alpha)}_{ij}}{x} \hat{n}_k\right)}\mathcal{P}\right]\nonumber \\
		&&+ (-1)^\alpha\boldsymbol{\mathcal{R}}^\alpha \cdot \left[\bigg<\frac{\boldsymbol{\Lambda}^{(\alpha)} V_{\beta}}{x}\bigg> \cdot(\mathbf{1}-\nn\nn)\cdot \uu^\beta \mathcal{P}\right]\nonumber\\
		&& - \kB T (\nn\times \rotop)_k(\mathbf{1}-\nn\nn)_{ik}(-1)^\alpha\left[\mathcal{R}^{\alpha}_{j}\left(\moy{\frac{\Lambda^{(\alpha)}_{ij}}{x}}\mathcal{P}\right)- \moy{\mathcal{R}^\alpha_{j}\frac{\Lambda^{(\alpha)}_{ij}}{x}}\mathcal{P}\right]\nonumber\\
		&& + (\nn\times \rotop)\cdot (\mathbf{1}-\nn\nn)\cdot\left[\bigg<\frac{\boldsymbol{\Lambda}^{(\alpha)} V_{12}}{x}\bigg> \cdot (\uu^\alpha \times \uu^\beta)\mathcal{P} + \bigg<\frac{\boldsymbol{\Lambda}^{(\alpha)} V_{\alpha}}{x}\bigg> \cdot (\uu^\alpha \times \nn)\mathcal{P}\right]\Bigg\}
		\label{LRPav}
		\end{eqnarray}
		We have used the relation 
		\begin{equation}
		\moy{U' \phi(x)} = \kB T \moy{\phi'(x)+2\, \frac{\phi(x)}{x}},
		\label{equilibrium_condition}
		\end{equation}
		valid for any function $\phi$ of the separation coordinate and under the assumption that the $x$-dependence of $P$ is Boltzmann-like, so that $P\propto \ex{-U/\kB T}$. Note that there are no terms with the coefficient $V_{12}$ in  (\ref{LTPav}), as such terms are due to the action of rotation gradient operator $\rotop^\alpha$ on the interaction potential.
	 \section{Moment expansion}
The next step in the calculation is a treatment of the orientations of the dumbbell. Analytic studies of low Reynolds number dynamics typically involve a description of the continuum equations; following previous work \cite{saha_clusters_2014,ahmadi_hydrodynamics_2006,marchetti_hydrodynamics_2013,saintillan_instabilities_2008} we turn to a macroscopic description of the system where we consider the evolution of the probability density $\rho=\int_{\nn,\uu^1,\uu^2}\mathcal{P}$. The local density $\rho$, and the local polarisations $\pp$, $\pp^\alpha$, are constant over time-scales that large compared to the time-scale of fluctuations, and lengths that are are large on the scale of the typical size of the dumbbell, and so are appropriate quantities for describing diffusion of the centre-of-mass \cite{ahmadi_hydrodynamics_2006}. However, the evolution equation for the density is not closed and involves polarisation fields $\pp=\int_{\nn,\uu^1,\uu^2}\nn\mathcal{P}$ and $\pp^\alpha=\int_{\nn,\uu^1,\uu^2}\uu^\alpha\mathcal{P}$---inevitably, the equations satisfied by the polarisation fields are also not closed and involve higher-order moments of the distribution such as the second order nematic tensor $\mathbf{Q}^{\alpha \beta}= \int \left(\uu^{\alpha} \uu^{\beta}- \frac{1}{3}\delta^{\alpha \beta}\mathbf{1}\right)\mathcal{P}$ (where $\alpha,\beta=0$ corresponds to $\nn$), which must be re-expressed in terms of lower order moments.\par
		A discussion of orientation requires a statement on the functional dependence of the hydrodynamic tensors on the orientations of the dumbbell. With the orientation-pre-averaging approximation, where hydrodynamic tensors were approximated as isotropic, and written $\mathbf{A}\simeq a_0\mathbf{1}$, we had previously described a coarse-grained dynamics. We now switch on anisotropic hydrodynamic interactions.\par
		The hydrodynamic tensors in Eqs. (\ref{LTP}) and (\ref{LRP}) are functions of the geometry, orientation, and separation of the subunits. The dependence on separation at least will be non-linear, but an explicit form of this dependence is not required for our analysis so we make no further speculation here. Rather, we posit the following expansion in orientations:
		\begin{eqnarray}
		\label{mob_expansion}
		\mathbf{M}& = & \mathbf{M}^{11}_{\mbox{\scriptsize TT}} + \mathbf{M}^{22}_{\mbox{\scriptsize TT}} + 2\mathbf{M}^{12}_{\mbox{\scriptsize TT}} \simeq [m_0(x) + m_1(x)\uu^1 \cdot \nn + m_2(x)\uu^2 \cdot \nn + m_{12}(x)\uu^1 \cdot \uu^2 + \dots] \; \mathbf{1}, \nonumber \\
		\mathbf{W}& = & \mathbf{M}^{11}_{\mbox{\scriptsize TT}} + \mathbf{M}^{22}_{\mbox{\scriptsize TT}} - 2\mathbf{M}^{12}_{\mbox{\scriptsize TT}} \simeq[w_0(x) + w_1(x)\uu^1 \cdot \nn + w_2(x)\uu^2 \cdot \nn + w_{12}(x)\uu^1 \cdot \uu^2 + \dots] \; \mathbf{1}, \nonumber\\
		\boldsymbol{\Gamma}& = & \mathbf{M}^{22}_{\mbox{\scriptsize TT}} - \mathbf{M}^{11}_{\mbox{\scriptsize TT}}\simeq [\gamma_0(x) + \gamma_1(x)\uu^1 \cdot \nn + \gamma_2(x)\uu^2 \cdot \nn + \gamma_{12}(x)\uu^1 \cdot \uu^2 + \dots] \; \mathbf{1},\nonumber\\
		\boldsymbol{\Phi}^{(1)}& = & \mathbf{M}^{11}_{\mbox{\scriptsize TR}}+\mathbf{M}^{21}_{\mbox{\scriptsize TR}}\simeq [\phi^{(1)}_0(x) + \phi^{(1)}_1(x)\uu^1 \cdot \nn + \phi^{(1)}_2(x)\uu^2 \cdot \nn + \phi^{(1)}_{12}(x)\uu^1 \cdot \uu^2 + \dots] \; \mathbf{1},\nonumber\\
		\boldsymbol{\Phi}^{(2)}& = & \mathbf{M}^{22}_{\mbox{\scriptsize TR}}+\mathbf{M}^{12}_{\mbox{\scriptsize TR}}\simeq[\phi^{(2)}_0(x) + \phi^{(2)}_1(x)\uu^1 \cdot \nn + \phi^{(2)}_2(x)\uu^2 \cdot \nn + \phi^{(2)}_{12}(x)\uu^1 \cdot \uu^2 + \dots]\;  \mathbf{1},\nonumber \\
		\boldsymbol{\Lambda}^{(1)}& = & \mathbf{M}^{11}_{\mbox{\scriptsize TR}}-\mathbf{M}^{21}_{\mbox{\scriptsize TR}}\simeq [\lambda^{(1)}_0(x) + \lambda^{(1)}_1(x)\uu^1 \cdot \nn + \lambda^{(1)}_2(x)\uu^2 \cdot \nn + \lambda^{(1)}_{12}(x)\uu^1 \cdot \uu^2 + \dots] \;  \mathbf{1},\nonumber \\
		\boldsymbol{\Lambda}^{(2)}& = & \mathbf{M}^{22}_{\mbox{\scriptsize TR}}-\mathbf{M}^{12}_{\mbox{\scriptsize TR}}\simeq [\lambda^{(2)}_0(x) + \lambda^{(2)}_1(x)\uu^1 \cdot \nn + \lambda^{(2)}_2(x)\uu^2 \cdot \nn + \lambda^{(2)}_{12}(x)\uu^1 \cdot \uu^2 + \dots] \; \mathbf{1},\nonumber \\
		\boldsymbol{\Psi}^{(1)}& = & \mathbf{M}^{11}_{\mbox{\scriptsize RR}}\simeq [\psi^{(1)}_0(x) + \psi^{(1)}_1(x)\uu^1 \cdot \nn + \psi^{(1)}_2(x)\uu^2 \cdot \nn + \psi^{(1)}_{12}(x)\uu^1 \cdot \uu^2 + \dots] \; \mathbf{1},\nonumber \\
		\boldsymbol{\Psi}^{(2)}& = & \mathbf{M}^{22}_{\mbox{\scriptsize RR}}\simeq[\psi^{(2)}_0(x) + \psi^{(2)}_1(x)\uu^1 \cdot \nn + \psi^{(2)}_2(x)\uu^2 \cdot \nn + \psi^{(2)}_{12}(x)\uu^1 \cdot \uu^2 + \dots] \; \mathbf{1},\nonumber\\
		\boldsymbol{X}^{(\alpha \beta)}& = & \mathbf{M}^{\alpha \beta}_{\mbox{\scriptsize RR}}\simeq[\chi^{(\alpha \beta)}_0(x) + \chi^{(\alpha \beta)}_1(x)\uu^1 \cdot \nn + \chi^{(\alpha \beta)}_2(x)\uu^2 \cdot \nn + \chi^{(\alpha \beta)}_{12}(x)\uu^1 \cdot \uu^2 + \dots] \; \mathbf{1},
		\end{eqnarray}
		again to first order in the angles of rotation. In this notation coefficients with a subscript zero are order zero in the moment expansion and otherwise first order.\par 
		With the inclusion of anisotropy in the hydrodynamic tensors, the zeroth and first order moments of the separation-averaged Smoluchowski equation give the following dynamical equations for the density and polarisation fields: 
		\begin{eqnarray}
		\partial_t\rho & = & \frac{\kB T}{4} \moy{m_0}\nabla^2_{\RR}\rho + \kB T \moy{\frac{\gamma_0}{x}} \nabla_{\RR} \cdot \pp + \sum_{\alpha\neq\beta}\Bigg\{\frac{1}{3}\moy{\frac{\gamma_0 V_{\alpha}}{x}}\nabla_{\RR}\cdot\pp^\alpha + \frac{1}{9}\moy{\frac{\gamma_{12} V_{\alpha}}{x}}\nabla_{\RR}\cdot\pp^\beta\Bigg\}, 
		\label{rhoT}\\
		\partial_t p_i & = & -\frac{\kB T}{3} \moy{\frac{\gamma_0}{x}} \partial_{R_i}\rho -  2\kB T \moy{\frac{w_0}{x^2}}p_i - \frac{2}{3}\sum_{\alpha\neq\beta}\Bigg\{\moy{\frac{w_0V_\alpha}{x^2}}p^\alpha_i + \frac{1}{3}\moy{\frac{w_{12}V_\alpha}{x^2}}p_i^\beta\Bigg\},
		\label{pT}\\
		\partial_t p^\alpha_{i} & = & \frac{1}{9}\left[\moy{\frac{\gamma_{0} V_{\alpha}}{x}} + \frac{1}{3}\moy{\frac{\gamma_{12} V_{\beta}}{x}}\right]\partial_{R_i}\rho -\frac{2}{3}\left[\moy{\psi_{0}^{(\alpha)} V_{\alpha}} + \frac{1}{3}\moy{(\psi_{\beta}^{(\alpha)} - \chi_\beta^{(\alpha\beta)})V_{12}} + \frac{1}{3}\moy{\chi_{12}^{(\alpha\beta)}V_{\beta}}\right]p_i\nonumber \\
		&& -2\kB T\moy{\psi_0^{(\alpha)}}p_i^\alpha -\frac{2}{3}\left[\moy{(\psi_{0}^{(\alpha)}-\chi_0^{(\alpha\beta)})V_{12}} - \frac{1}{3}\moy{\chi_\alpha^{(\alpha\beta)}V_{\beta}} + \kB T\moy{\chi_{12}^{(\alpha\beta)}}\right]p^\beta_{i}.
		\label{palphaT}
		\end{eqnarray}
We note that the hydrodynamic equations are not closed because of asymmetry, and that the complications in the moment expansion are precisely due to this reason. The hierarchy is truncated by a closure scheme which yields a closed set of dynamical equations for the density and polarisation fields. In equations (\ref{rhoT})-(\ref{palphaT}), we have kept terms upto the lowest order in spatial derivatives ($\nabla^2_{\RR}\rho$, $\nabla_{\RR}\cdot\pp$ and $\nabla_{\RR}\cdot\pp^\alpha$), and rewritten higher-order moments according to the expressions in \ref{TST}, keeping only the traceless symmetric part of the corresponding tensors.
\begin{figure}
\centering
\includegraphics[width=0.6\textwidth]{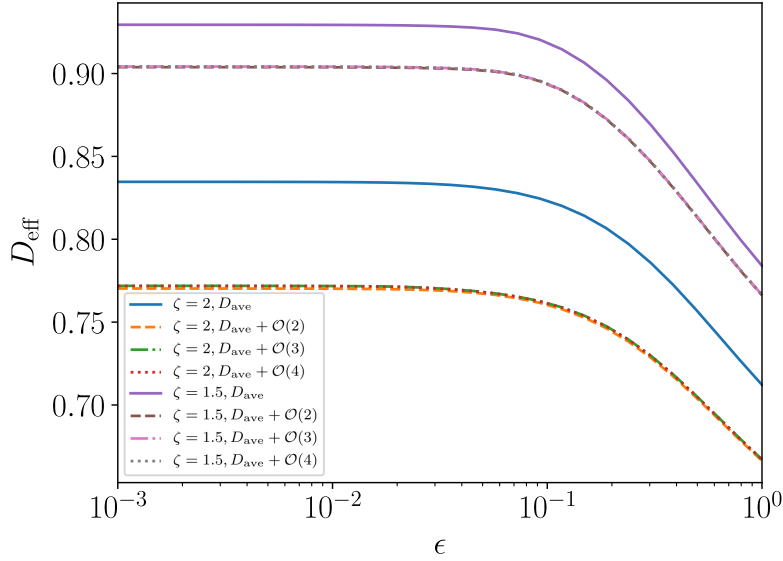}
\caption{The effective diffusion coefficient of two axisymmetric dumbbells with asymmetries $\zeta=2$ and $\zeta=1.5$ as a function of relative fluctuations $\ep$, with $v_1=0.1$, $v_2=0.13$, $v_{12}=0.15$ and $a_1/a=0.3$. $D_{\mbox{\scriptsize ave}}$ is the first term of $D_{\mbox{\scriptsize eff}}$; $D_{\mbox{\scriptsize ave}}+\mathcal{O}(2)$ is the sum of the first two terms; $D_{\mbox{\scriptsize ave}}+\mathcal{O}(3)$ includes $\delta D^{\alpha\alpha}_2$; and $D_{\mbox{\scriptsize ave}}+\mathcal{O}(4)$ includes $\delta D^{\alpha\beta}_3$.}
\label{fig:Deff-ep_main}
\end{figure}
\begin{figure}
\centering
\includegraphics[width=0.45\textwidth]{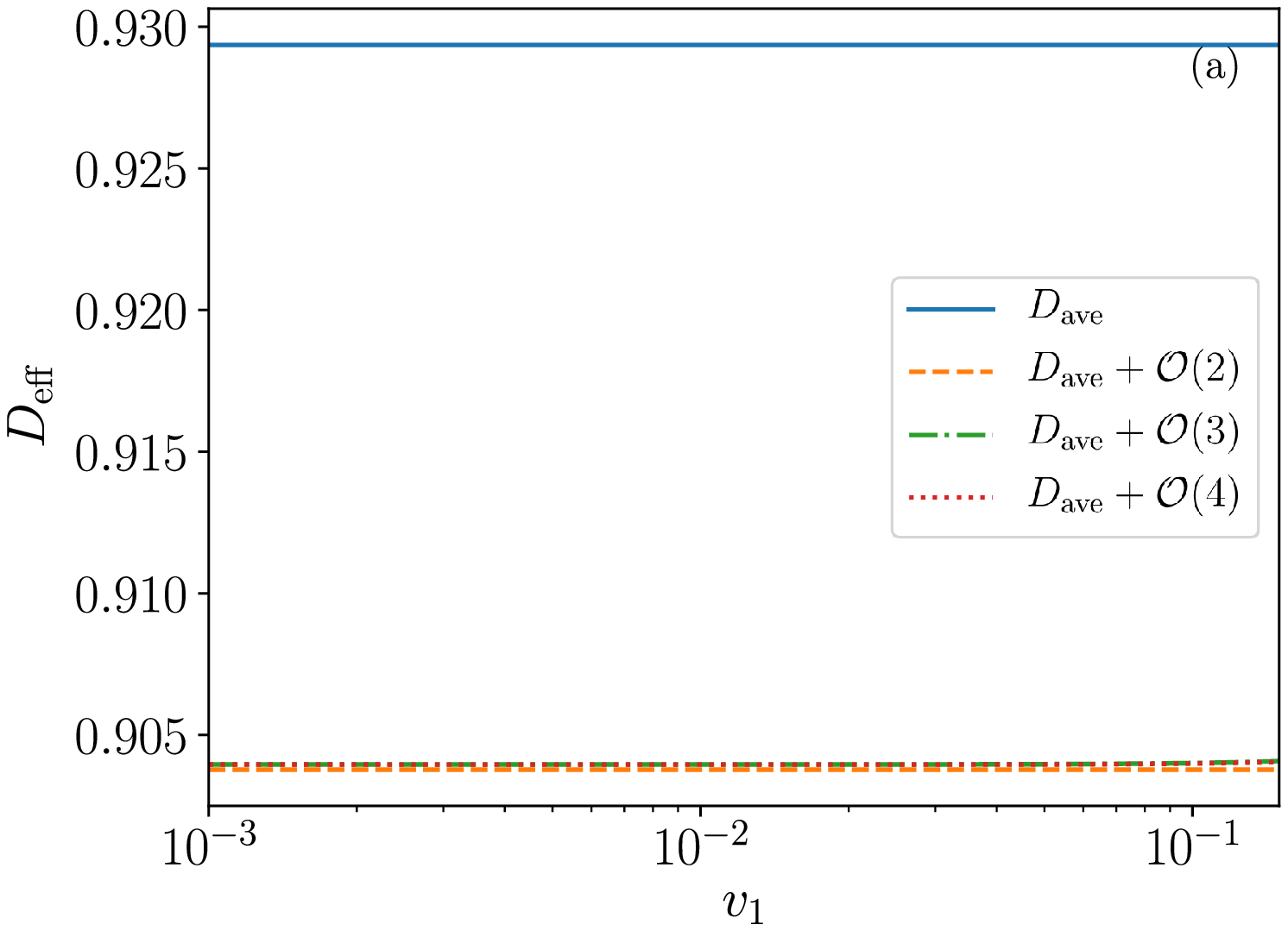}
\includegraphics[width=0.45\textwidth]{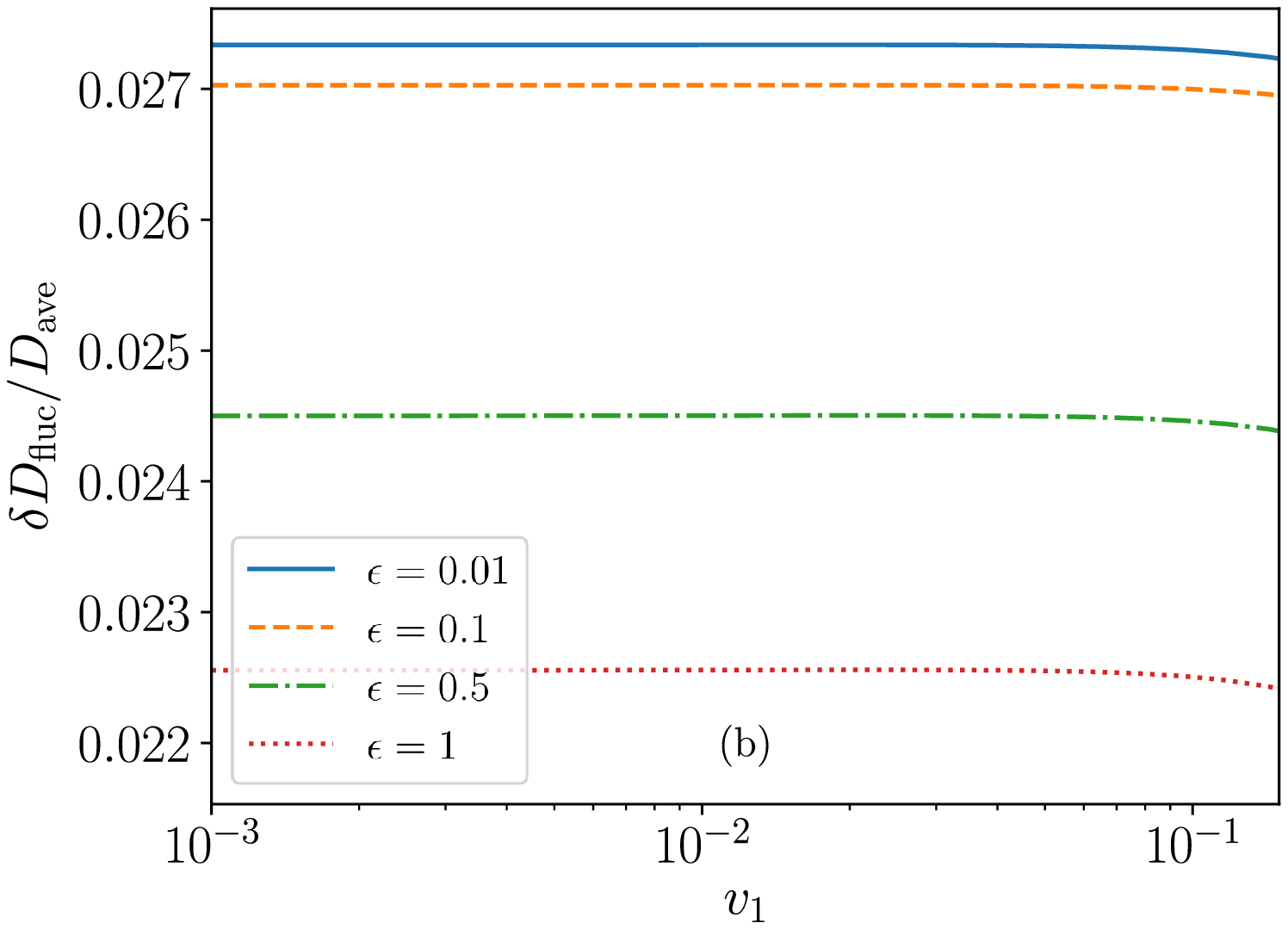}
\caption{In both figures, $\zeta=1.5$, $v_2=0.1$, $v_{12}=0.15$ and $a_1/a=0.3$. (a): The effective diffusion coefficient of an axisymmetric dumbbell as a function of the constant constraint parameter $v_1$, with $\epsilon=0.01$. (b): Relative decrease in diffusion coefficient for different values of $\epsilon$.}
\label{fig:Deff-v1}
\end{figure}
		\section{Closure approximation and the diffusion equation}
		\subsection{Closure scheme and its implications for alignment}
		From the set of equations (\ref{rhoT})-(\ref{palphaT}) we derive a closed diffusion equation with an appropriate closure scheme. By assuming time scales that are large compared to the relaxation time of fluctuations, we can assume $\partial_t\pp=0$ and $\partial_t\pp^\alpha=0$ 
and find expressions for the polarisation fields that are linear in density:
\begin{equation}
\label{closed_p}
p_i = \left[S_0+S_2+S_3 + \mathcal{O}(4)\right]\partial_{R_i}\rho
\end{equation}
where 
\begin{eqnarray}
S_0 & = & -\frac{1}{6}\frac{\moy{\gamma_0/x}}{\moy{w_0/x^2}},\nonumber \\
S_2 & = & -\frac{1}{6}\frac{1}{(3\kB T)}\frac{1}{\moy{w_0/x^2}}\sum_{\alpha \neq \beta}g_\alpha\moy{\frac{w_0V_{\alpha}}{x^2}},\nonumber \\
S_3 & = & - \frac{1}{18}\frac{1}{(3\kB T)^2}\frac{1}{\moy{w_0/x^2}}\sum_{\alpha \neq \beta}\moy{\frac{w_0V_{\alpha}}{x^2}}\frac{\moy{\frac{\gamma_{12}V_\beta}{x}}+\frac{\moy{\gamma_0/x}}{\moy{w_0/x^2}}\left[\moy{(\psi_{\beta}^{(\alpha)} - \chi_\beta^{(\alpha\beta)})V_{12}}+\moy{\chi_{12}^{(\alpha\beta)}V_{\beta}}\right]}{\moy{\psi_0^{(\alpha)}}}\nonumber \\
&& + \frac{1}{6}\frac{1}{(3\kB T)^2}\frac{1}{\moy{w_0/x^2}}\sum_{\alpha \neq \beta}g_\beta\moy{\frac{w_0V_{\alpha}}{x^2}}\frac{\moy{\left(\psi_0^{(\alpha)}-\chi_0^{(\alpha\beta)}\right)V_{12}} + \kB T\moy{\chi_{12}^{(\alpha\beta)}}}{\moy{\psi_0^{(\alpha)}}}\nonumber\\
&&-\frac{1}{18}\frac{1}{(3\kB T)}\frac{1}{\moy{w_0/x^2}}\sum_{\alpha \neq \beta}g_\beta\moy{\frac{w_{12}V_{\alpha}}{x^2}},
\end{eqnarray}
and
\begin{equation}
\label{closed_p_alpha}
p_i^{\alpha} = \left[T_1+T_2+T_3 + \mathcal{O}(4)\right]\partial_{R_i}\rho
\end{equation}
where
\begin{eqnarray}
T_1 & = & \frac{1}{6}g_\alpha,\nonumber \\
T_2 & = & \frac{1}{18}\frac{1}{(3\kB T)}\frac{\moy{\frac{\gamma_{12}V_\beta}{x}}+\frac{\moy{\gamma_0/x}}{\moy{w_0/x^2}}\left[\moy{(\psi_{\beta}^{(\alpha)} - \chi_\beta^{(\alpha\beta)})V_{12}}+\moy{\chi_{12}^{(\alpha\beta)}V_{\beta}}\right]}{\moy{\psi_0^{(\alpha)}}}\nonumber\\
&& -\frac{1}{6}\frac{1}{(3\kB T)}\frac{\moy{(\psi_{0}^{(\alpha)}-\chi_0^{(\alpha\beta)})V_{12}} + \kB T\moy{\chi_{12}^{(\alpha\beta)}}}{\moy{\psi_0^{(\alpha)}}}g_\beta,\nonumber\\
T_3 & = & \frac{1}{6}\frac{1}{(3\kB T)^2}\frac{1}{\moy{w_0/x^2}}\frac{\moy{\psi_0^{(\alpha)}V_\alpha}}{\moy{\psi_0^{(\alpha)}}}\sum_{\gamma}\moy{\frac{w_0V_{\gamma}}{x^2}}g_\gamma.
\end{eqnarray}
The dimensionless coefficient 
\begin{equation}
g_\alpha = \frac{1}{3\kB T\moy{\psi_0^{(\alpha)}}}\left[\moy{\frac{\gamma_0V_\alpha}{x}} + \frac{\moy{\gamma_0/x}}{\moy{w_0/x^2}}\moy{\psi_0^{(\alpha)}V_\alpha}\right]
\label{g_alpha}
\end{equation}
gives the scale of polarisation of subunit $\alpha$ at leading order: The scale of polarisation of the complex is given by $\moy{\gamma_0/x}/\moy{w_0/x^2}$ at leading order. The index of the coefficients $S_i$ and $T_i$ gives their order in the joint expansion of potential and mobilities. The effect of anisotropy is seen in the coefficients $S_3$ and $T_2$, that is, at second order in the polarisation of a subunit and at third order in the polarisation of the complex.
		\subsection{The diffusion equation}
		\begin{figure}
\centering
\includegraphics[width=0.6\textwidth]{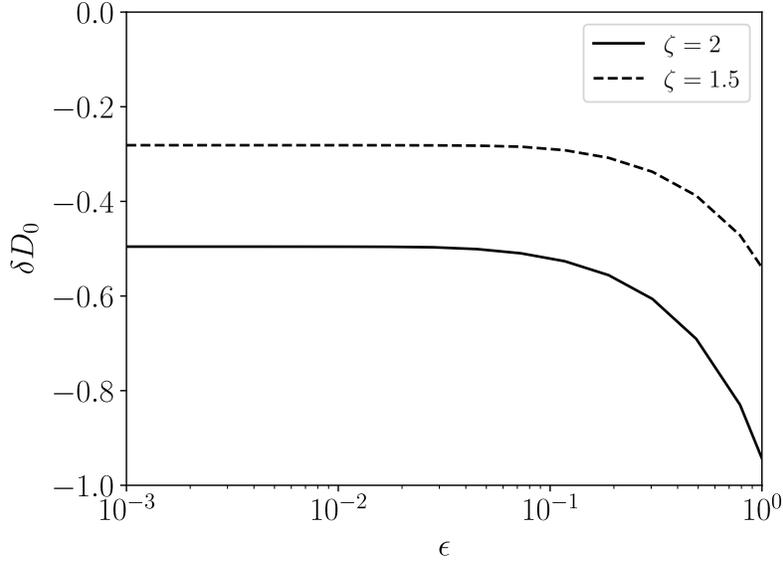}
\caption{$\delta D_0$ as a function of $\ep$, for $a_1/a=0.3$. $\delta D_0$ is the first term of $\delta D_{\mbox{\scriptsize fluc}}$, which is expected to be negative. The magnitude of $\delta D_0$ increases with asymmetry of the dumbbell.}
\label{fig:deltaD_0-ep}
\end{figure}
		In this limit the following effective diffusion coefficient is deduced 
		\begin{equation}
		\label{D_eff}
		D_{\mbox{\scriptsize eff}} = \frac{\kB T}{4}\moy{m_0}-\frac{\kB T}{6}\frac{\moy{\gamma_0/x}^2}{\moy{w_0/x^2}}\left[1-\sum_{\alpha\neq\beta}\left(\delta D^{\alpha\alpha}_2+\delta D^{\alpha\beta}_3\right)\right]+\mathcal{O}(4),
		\end{equation}
		where the first term is the thermal average of contributions from translational modes of the dumbbell; the leading order correction is due to asymmetry in the self mobilities of the subunits; and $\delta D^{\alpha\beta}_i$ is a fluctuation-induced correction of order $i$ in the moment expansion. Explicitly  
		\begin{eqnarray}
		\delta D^{\alpha\alpha}_{2} & = & \frac{\moy{w_0/x^2}}{\moy{\gamma_0/x}}h_\alpha g_\alpha, \nonumber\\
		\delta D^{\alpha\beta}_{3} & = & -\frac{1}{3\kB T}\frac{\moy{w_0/x^2}}{\moy{\gamma_0/x}}h_\alpha g_\beta\frac{\moy{\left(\psi_0^{(\alpha)}-\chi_0^{(\alpha\beta)}\right)V_{12}} + \kB T\moy{\chi_{12}^{(\alpha\beta)}}}{\moy{\psi_0^{(\alpha)}}}\nonumber\\
		&& + \frac{1}{3}\frac{1}{(3\kB T)}\frac{\moy{\gamma_0/x}}{\moy{w_0/x^2}}h_\alpha\frac{\left[\moy{(\psi_{\beta}^{(\alpha)} - \chi_\beta^{(\alpha\beta)})V_{12}}+\moy{\chi_{12}^{(\alpha\beta)}V_{\beta}}\right]\frac{\moy{w_0/x^2}}{\moy{\gamma_0/x}}+ \moy{\gamma_{12}V_\beta/x}}{\moy{\psi_0^{(\alpha)}}}\nonumber\\
		&& +\frac{1}{3}\frac{1}{(3\kB T)}\frac{1}{\moy{\gamma_0/x}}g_\alpha\left[\moy{\frac{w_{12}V_\beta}{x^2}} + \moy{\frac{\gamma_{12}V_\beta}{x}}\frac{\moy{\gamma_0/x}}{\moy{w_0/x^2}}\right],
		\end{eqnarray}
where $g_\alpha$ is as in (\ref{g_alpha}), and we have defined another dimensionless quantity  
\begin{equation}
h_\alpha = \frac{1}{3\kB T}\left[\frac{\moy{\gamma_0V_\alpha/x}}{\moy{\gamma_0/x}}-\frac{\moy{w_0V_\alpha/x^2}}{\moy{w_0/x^2}}\right].
\label{h_alpha}
\end{equation}	
Unlike $g_\alpha$, $h_\alpha$ does not vanish if the subunits are identical and is independent of rotational motion.\par
		 The motion of the dumbbell is retarded by asymmetry through hydrodynamic interactions between the subunits. The correction $\delta D^{\alpha\alpha}_2$, previously reported in \cite{illien_diffusion_2017}, is of order one and is estimated to be positive for harmonic-like potentials. 
Corrections are observed at a time-scale governed by the separation relaxation time (a detailed calculation is given in \cite{illien_diffusion_2017}). As expected from the equation for the rotational gradient of the interaction potential (\ref{rotop_U}), $V_{12}(x)$ is coupled to the hydrodynamic tensors for rotational modes of the dumbbell. The effect of anisotropy is first seen at third order where it is coupled to asymmetry. A notable absence in $D_{\mbox{\scriptsize eff}}$ is that of translation-rotation coupling which would be introduced by the tensors $\boldsymbol{\Phi}^{(\alpha)}$ and $\boldsymbol{\Lambda}^{(\alpha)}$, representing the off-diagonal blocks of the mobility matrix. Such a coupling will be produced if the system is given a feature which breaks the corresponding chiral symmetry, such as a macroscopic constraint  \cite{goldfriend_hydrodynamic_2016}.\par
		Having truncated the moment expansion at third order, eq. (\ref{D_eff}) for the effective diffusion coefficient holds for sufficiently small constraint functions $V_\alpha(x)$ and $V_{12}(x)$. Within the regime considered, we suggest that additional terms which are the result of higher order contributions in either the potential or mobility expansions will have only minor quantitative contributions to $D_{\mbox{\scriptsize eff}}$. In the same way, we expect that fluctuations about the two remaining Euler angles of each subunit yield separate, but comparable corrections---overall contributing negatively.\par
		To fourth order in the moment expansion, we can write the fluctuation-induced corrections as the product of internal ($\moy{\gamma_0/x}$) and external asymmetry by comparison with (\ref{closed_p}) and (\ref{closed_p_alpha})
		\begin{eqnarray}
\delta D_{\mbox{\scriptsize fluc}} & = &\frac{\kB T}{6}\moy{\frac{\gamma_0}{x}}\Bigg[\frac{\moy{\gamma_0/x}}{\moy{w_0/x^2}} - h_\alpha g_\alpha + h_\alpha g_\beta\frac{\moy{\left(\psi_0^{(\alpha)}-\chi_0^{(\alpha\beta)}\right)V_{12}} + \kB T\moy{\chi_{12}^{(\alpha\beta)}}}{3\kB T\moy{\psi_0^{(\alpha)}}}\nonumber\\
&& - \frac{\moy{\gamma_0/x}}{\moy{w_0/x^2}}h_\alpha\frac{\left[\moy{(\psi_{\beta}^{(\alpha)} - \chi_\beta^{(\alpha\beta)})V_{12}}+\moy{\chi_{12}^{(\alpha\beta)}V_{\beta}}\right]+ \moy{\gamma_{12}V_\beta/x}\frac{\moy{\gamma_0/x}}{\moy{w_0/x^2}}}{9\kB T\moy{\psi_0^{(\alpha)}}}\nonumber\\
&& - \frac{1}{9\kB T}\frac{1}{\moy{w_0/x^2}}g_\alpha\left[\moy{\frac{w_{12}V_\beta}{x^2}} + \moy{\frac{\gamma_{12}V_\beta}{x}}\frac{\moy{\gamma_0/x}}{\moy{w_0/x^2}}\right]\Bigg].
\label{new_factorisation}
\end{eqnarray}	
Simply, since the reduction in diffusion coefficient is driven by asymmetry, any reduction in asymmetry will lead to a positive contribution to $D_{\mbox{\scriptsize eff}}$. If the external symmetry is broken the effect of symmetry breaking on the diffusion coefficient can be determined by recalculating the polarisation fields with an appropriate expansion to replace (\ref{mob_expansion}). The equilibrium model proposed in Ref. \cite{illien_diffusion_2017} for a catalytically active enzyme demonstrates the case when there is a change in internal symmetry. In the case of internal symmetry breaking the effect is introduced by redefining the average over the internal degrees of freedom.
\begin{figure}
\centering
\includegraphics[width=0.45\textwidth]{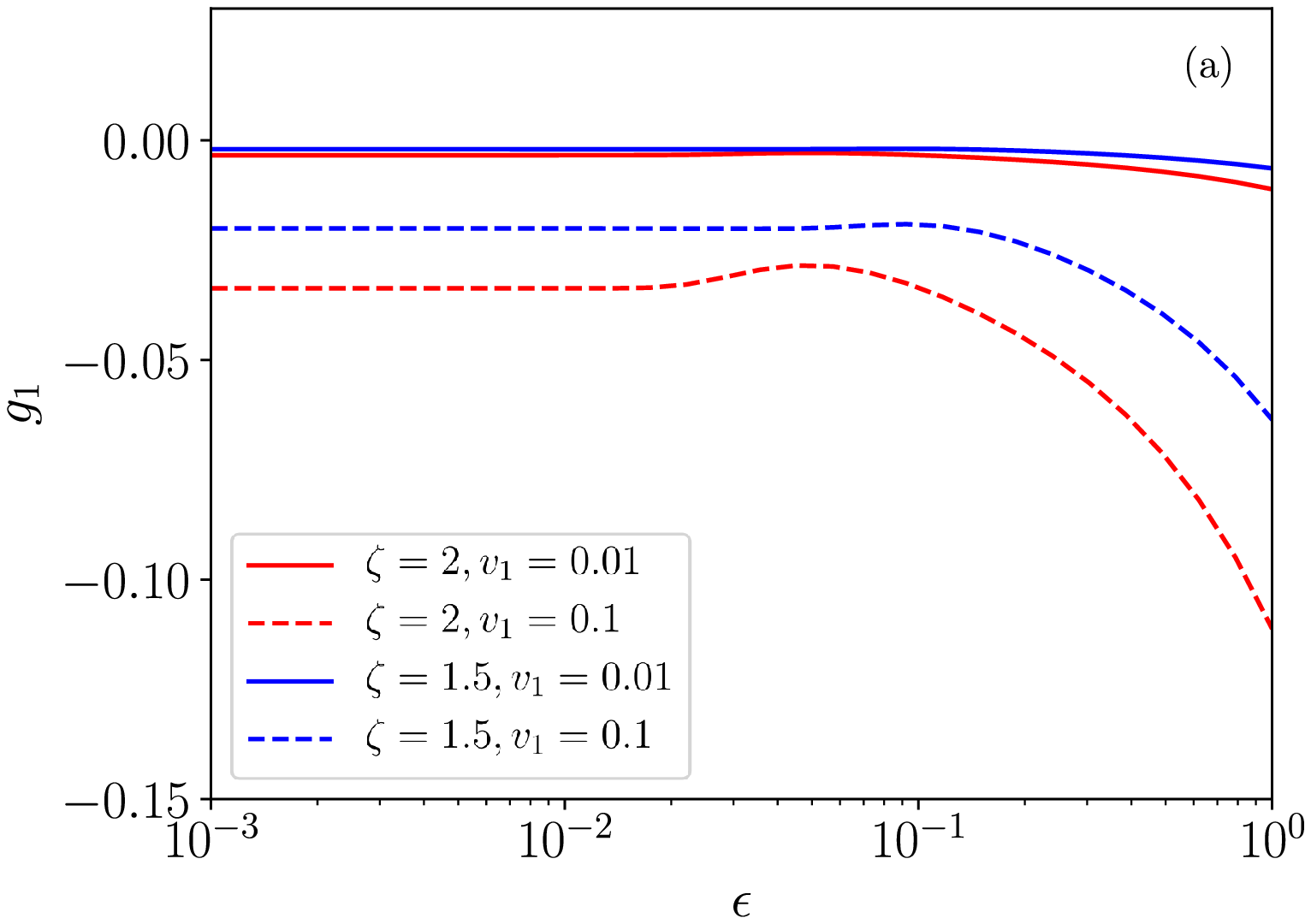}
\includegraphics[width=0.45\textwidth]{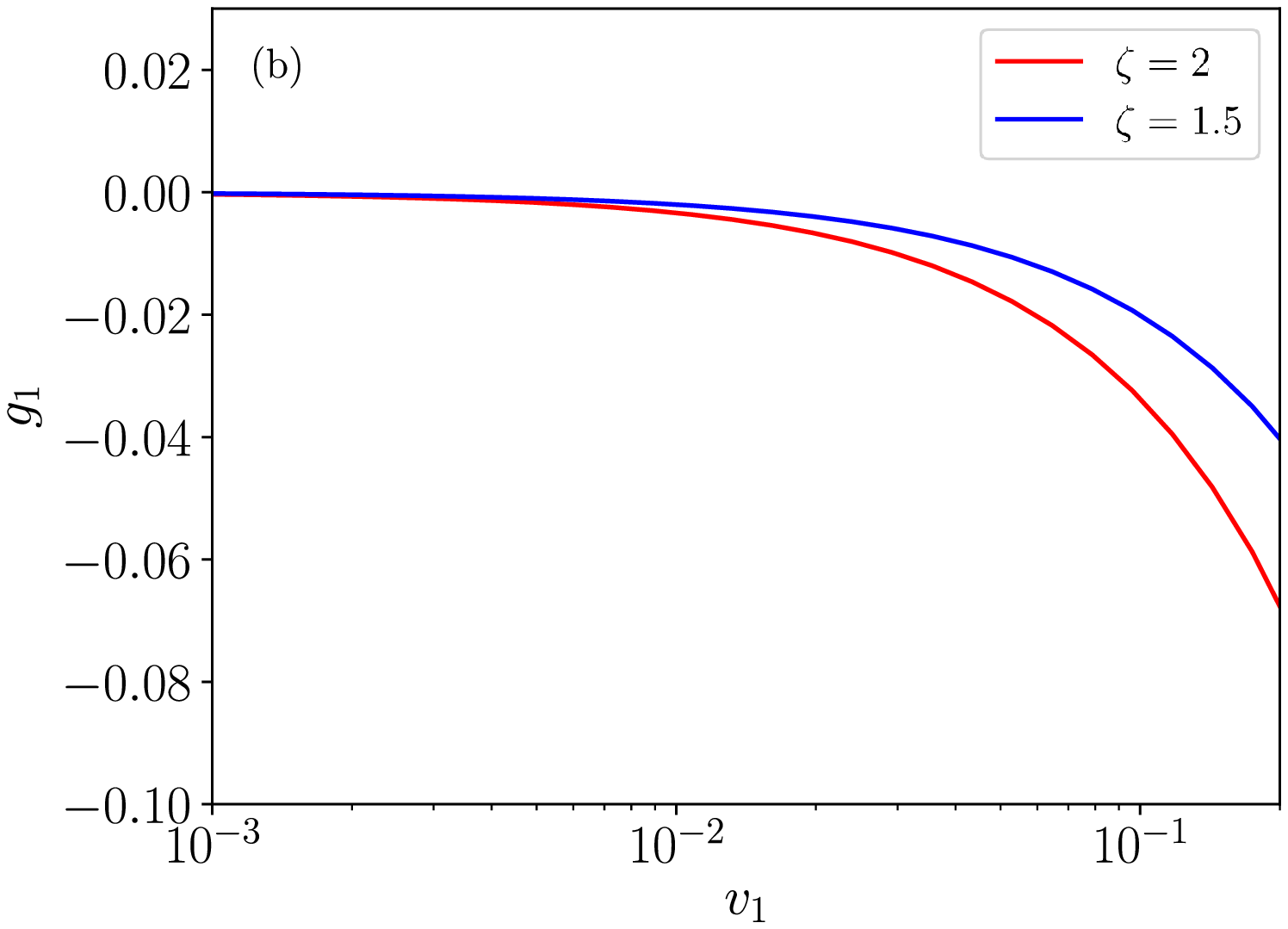}
\caption{(a): The coefficient $g_1$ as a function of $\ep$. (b): $g_1$ as a function of $v_1 \ll 1$, with $a_1/a=0.3$.}
\label{fig:g_1}
\end{figure}
\section{Harmonic potential}
To provide a more quantitative analysis of the role of the interaction on $D_{\mbox{\scriptsize eff}}$, we consider an example of rigid spherical subunits of radii $a_1$ and $a_2$ with $a_2>a_1$. The subunits interact via a potential $U$ that has harmonic $x$-dependence so that $U=\frac{1}{2}k(x-a)^2\left[1+\sum_{\alpha=1,2}v_\alpha\,\nn\cdot\uu^\alpha + v_{12}\,\uu^1\cdot\uu^2\right]$ for $x>a_1+a_2$ and $\infty$ otherwise. In $U$, $a$ is the typical size of the dumbbell and $v_\alpha$ and $v_{12}$ are the constraint parameters which are independent of $x$. The mobility functions for interacting particles are widely known for such axisymmetric geometries. We provide some preliminary plots of $D_{\mbox{\scriptsize eff}}$, which were made using the mobility functions of \cite{jeffrey_calculation_1984} for widely separated spheres.
We have defined the dimensionless numbers $\ep=\sqrt{\kB T/ka^2}$, a measure of the fluctuations of the dumbbell around its equilibrium, $\zeta=a_2/a_1$, the geometric asymmetry and the ratio $a_1/a$. Since we perform an expansion in the constraints, (\ref{D_eff}) is bounded by $v_\alpha, v_{12} \ll 1$.\par
Fig. \ref{fig:Deff-ep_main} shows the relation between $D_{\mbox{\scriptsize eff}}$ and the relative fluctuations of the dumbbell. As expected, the diffusion coefficient of the dumbbell is lowered when its asymmetry is increased. Furthermore, $D_{\mbox{\scriptsize eff}}$ increases with the stiffness of the potential, and its maximum value is attained in the limit of a rigid potential.\par
Fig. \ref{fig:Deff-v1}(a) shows the dependence of $D_{\mbox{\scriptsize eff}}$ on the constraint parameter $v_1$, with constant $v_2$ and $v_{12}$. There is a clear reduction in the effective diffusion upon the inclusion of the fluctuation induced corrections, particularly $\delta D^{\alpha\alpha}_2$, even at small relative deformations. In the far-field limit where the separation of the subunits is much greater than the typical size of the dumbbell, the mobility functions are given by a series expansion in $1/x$ so that a small deformation in the dumbbell can have significant contributions to the diffusion coefficient. In fig. \ref{fig:Deff-v1}(b) it can be seen that the relative decrease in the diffusion coefficient due to fluctuation-induced corrections increases with the stiffness of the potential.\par
\begin{figure}
   \centering
        \includegraphics[width=0.45\textwidth]{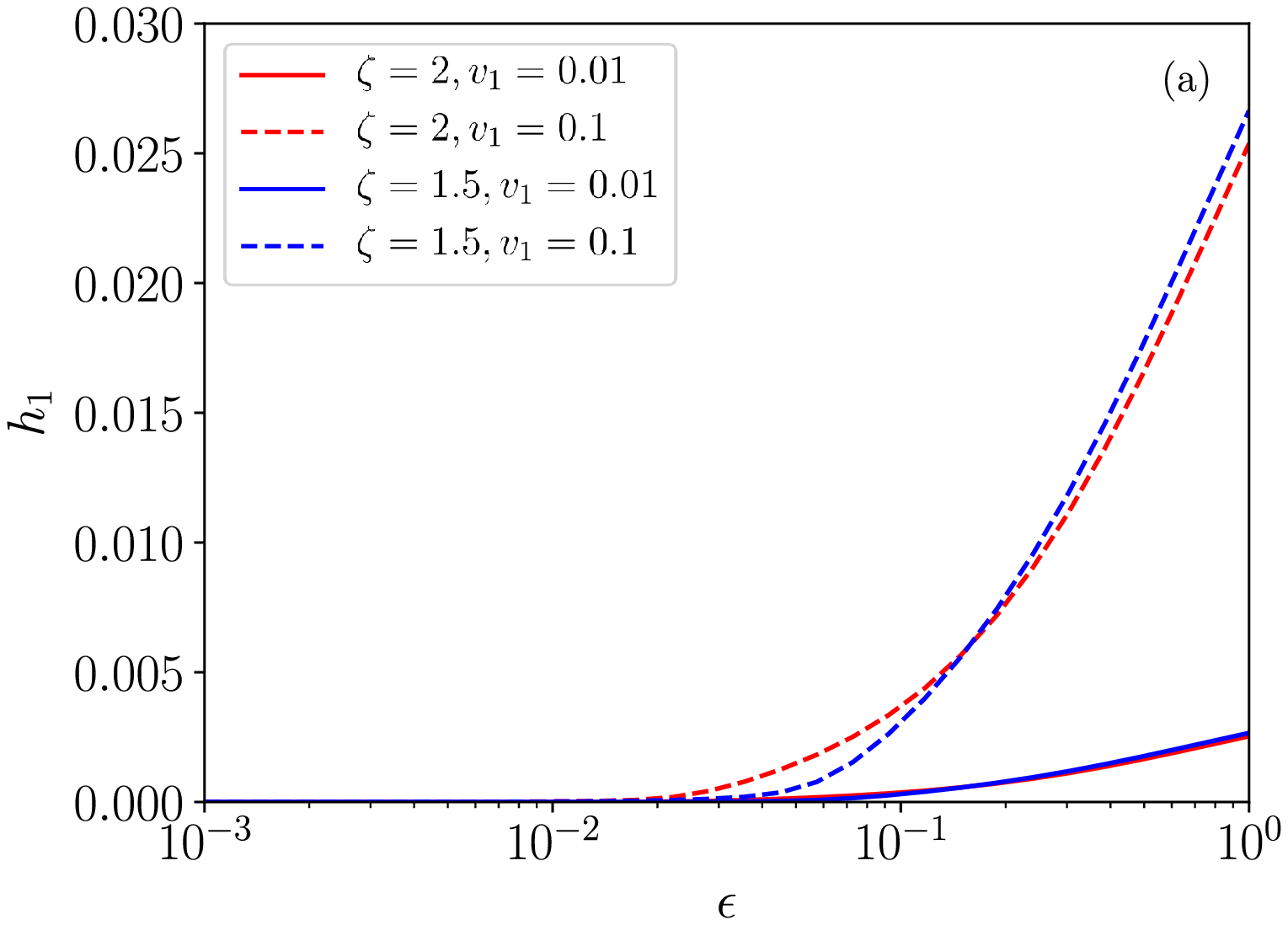}
        \includegraphics[width=0.45\textwidth]{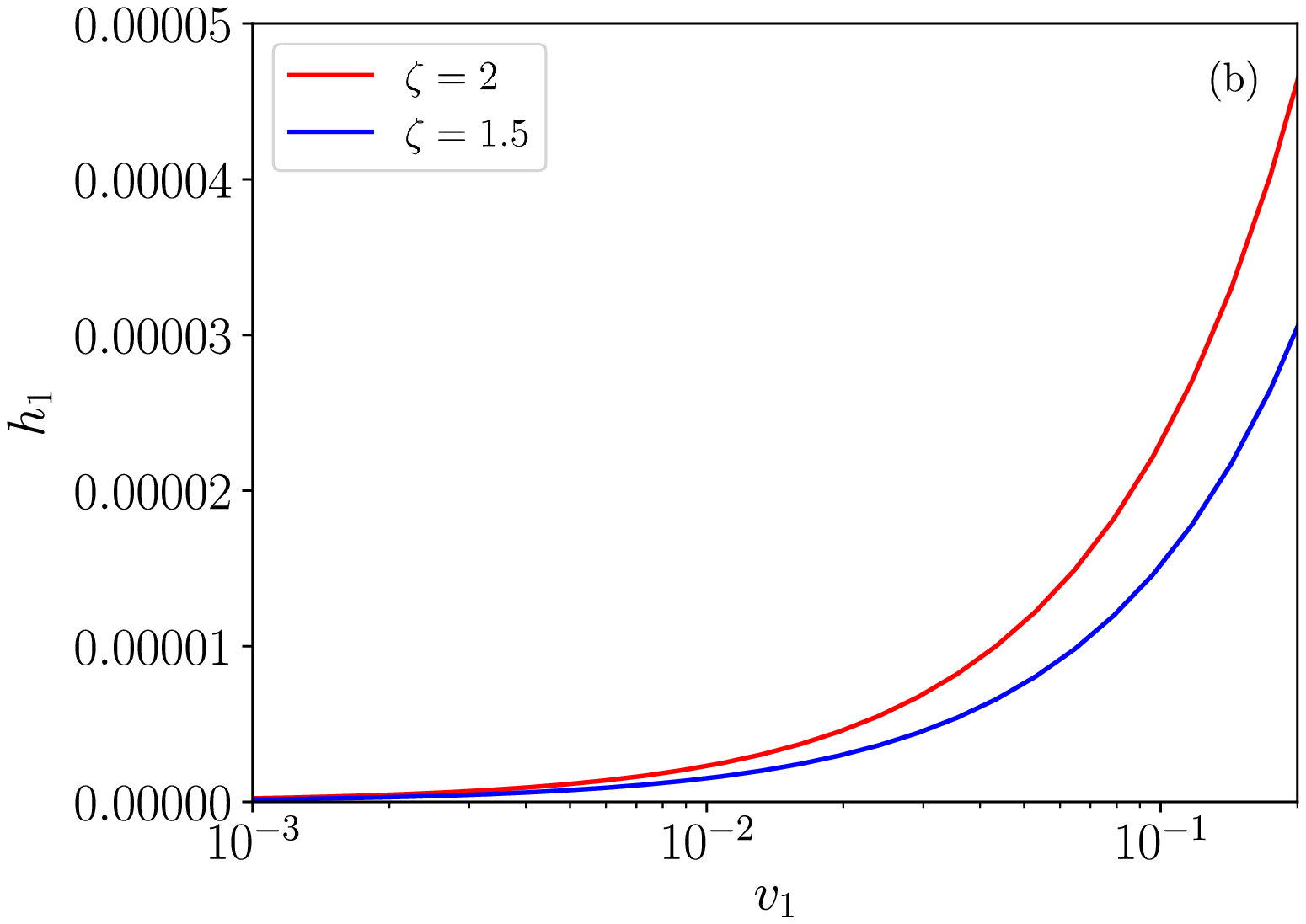}
        \caption{(a): The coefficient $h_1$ as a function of $\ep$. (b): $h_1$ as a function of $v_1 \ll 1$, with $a_1/a=0.3$.}
\label{fig:h_1}
\end{figure}

It is instructive to study the behaviour of the coefficients $\delta D_0=\moy{\gamma_0/x}/\moy{w_0/x^2}$, $g_\alpha$ and $h_\alpha$ which are present in every term of $\delta D_{\mbox{\scriptsize fluc}}$ and are fully captured by an axisymmetric geometry. In figures \ref{fig:deltaD_0-ep}-\ref{fig:h_1} we show their dependence on the constraint parameter $v_1$ and relative fluctuations for two axisymmetric dumbbells. It can be seen that $\delta D_0$ has the largest contribution which is negative; typical values of $g_\alpha$ are negative but an order of magnitude smaller; and as a function of $v_\alpha$ and $\ep$, $h_\alpha$ is positive, but crucially negligible. Therefore, within the regime of the moment expansion, terms linear in $h_\alpha$ are negligible in $\delta D_{\mbox{\scriptsize fluc}}$. The last surviving term in (\ref{new_factorisation}) cannot be determined in this example because it contains anisotropy. Nonetheless, it is expected to be considerably smaller than $\delta D_0$.
\section{Conclusion}
We have proposed a new theoretical framework for understanding the effect of internal hydrodynamic interactions on the stochastic translational motion of modular structures in a fluid environment under low Reynolds number conditions. In an asymmetric modular structure, such interactions (here caused by thermal fluctuations) result in a decrease in the diffusion coefficient of the object, hence there is a fluctuation--induced correction to the diffusion coefficient. We show that this correction is driven by an interplay between internal and external asymmetry. Although the full result is presented for a general dumbbell, where anisotropy is a relevant measure, we consider the case of an axisymmetric dumbbell (for which the hydrodynamic functions are well known) to provide a quantitative example.\par 
We have shown that the hydrodynamic equations which describe the evolution of moments of the distribution function of the generalised dumbbell are not closed, but with a careful consideration of the motion of such an object, one is able to close these equations at any order.\par
The present work then gives an insight into the effect of internal fluctuations and asymmetries on the effective diffusion coefficient of a model modular macromolecule. The ideas presented here are very generic, and could be specified to different real systems. In particular, it would be of interest to study more complex structures that reproduce faithfully the structure of specific enzyme molecules. This could be achieved using extensive numerical simulations that would include both stochastic dynamics and hydrodynamic interactions, and could be the topic of future work.
\ack
TAL acknowledges the support of EPSRC. PI and RG were supported by the US National Science Foundation under MRSEC Grant number DMR-1420620.
\section*{References}
\bibliography{references_njp}

\providecommand{\newblock}{}
\begin{thebibliography}{10}
\expandafter\ifx\csname url\endcsname\relax
  \def\url#1{{\tt #1}}\fi
\expandafter\ifx\csname urlprefix\endcsname\relax\def\urlprefix{URL }\fi
\providecommand{\eprint}[2][]{\url{#2}}

\bibitem{alberts_2014}
Alberts B, Johnson A, Lewis J, Morgan D, Raff M~C, Walter P and Roberts K 2014
  {\em Molecular Biology of the Cell\/} (New York: Garland Science)

\bibitem{muddana_2010}
Muddana H~S, Sengupta S, Mallouk T~E, Sen A and Butler P~J 2010 {\em J. Am.
  Chem. Soc.\/} {\bf 132} 2110--2111

\bibitem{sengupta_2014}
Sengupta S, Spiering M~M, Dey K~K, Duan W, Patra D, Butler P~J, Astumian D~R,
  Benkovic S~J and Sen A 2014 {\em ACS nano\/} {\bf 8} 2410--2418

\bibitem{sengupta_2013}
Sengupta S, Dey K~K, Muddana H~S, Tabouillot T, Ibele M~E, Butler P~J and Sen A
  2013 {\em J. Am. Chem. Soc.\/} {\bf 135} 1406--1414

\bibitem{riedel_2014}
Riedel C, Gabizon R, Wilson C~A~M, Hamadani K, Tsekouras K, Marqusee S, Pressé
  S and Bustamante C 2014 {\em Nature\/} {\bf 517} 227--230

\bibitem{switala_2002}
Switala J and Loewen P~C 2002 {\em Arch. Biochem. Biophys\/} {\bf 401} 145--154

\bibitem{golestanian_2015}
Golestanian R 2015 {\em Phys. Rev. Lett.\/} {\bf 115} 108102

\bibitem{mikhailov_2015}
Mikhailov A~S and Kapral R 2015 {\em Proc. Natl. Acad. Sci. U.S.A\/} {\bf 112}
  E3639--E3644

\bibitem{golestanian-ajdari_2008}
Golestanian R and Ajdari A 2008 {\em Phys. Rev. Lett.\/} {\bf 100}(3) 038101

\bibitem{cressman_2008}
Cressman A, Togashi Y, Mikhailov A~S and Kapral R 2008 {\em Phys. Rev. E\/}
  {\bf 77} 050901

\bibitem{bai_2015}
Bai X and Wolynes P~G 2015 {\em J. Chem. Phys.\/} {\bf 143} 165101

\bibitem{illien_exothermicity_2017}
Illien P, Zhao X, Dey K~K, Butler P~J, Sen A and Golestanian R 2017 {\em Nano
  Lett.\/} {\bf 17} 4415--4420

\bibitem{illien_diffusion_2017}
Illien P, Adeleke-Larodo T and Golestanian R 2017 {\em EPL\/} {\bf 119} 40002

\bibitem{doi_theory_1986}
Doi M and Edwards S~F 1986 {\em The Theory of Polymer Dynamics\/}
  (International series of monographs on physics vol 73) (Oxford: Clarendon
  Press)

\bibitem{zimm_dynamics_1956}
Zimm B~H 1956 {\em J. Chem. Phys.\/} {\bf 24} 269--278

\bibitem{ottinger_rouse_1989}
\"Ottinger H~C 1989 {\em J. Chem. Phys.\/} {\bf 90} 463--473

\bibitem{ottinger_dumbbell_1989}
\"Ottinger H~C 1989 {\em Colloid Polym. Sci.\/} {\bf 267} 1--8

\bibitem{wilemski_1974}
Wilemski G and Fixman M 1974 {\em J. Chem. Phys.\/} {\bf 60} 866--877

\bibitem{levernier_2015}
Levernier N, Dolgushev M, B\'eénichou O, Blumen A, Gu\'eérin T and Voituriez
  R 2015 {\em J. Chem. Phys.\/} {\bf 143} 204108

\bibitem{happel1973low}
Happel J and Brenner H 1973 {\em Low Reynolds Number Hydrodynamics\/} Mechanics
  of fluids and transport processes (Kluwer Academic)

\bibitem{kim_microhydrodynamics:_2005}
Kim S and Karrila S~J 2005 {\em Microhydrodynamics: Principles and Selected
  Applications\/} (Mineola N.Y.: Dover Publications)

\bibitem{saha_clusters_2014}
Saha S, Golestanian R and Ramaswamy S 2014 {\em Phys. Rev. E\/} {\bf 89}

\bibitem{ahmadi_hydrodynamics_2006}
Ahmadi A, Marchetti M~C and Liverpool T~B 2006 {\em Phys. Rev. E\/} {\bf 74}
  061913

\bibitem{marchetti_hydrodynamics_2013}
Marchetti M~C, Joanny J~F, Ramaswamy S, Liverpool T~B, Prost J, Rao M and Simha
  R~A 2013 {\em Rev. Mod. Phys.\/} {\bf 85} 1143--1189

\bibitem{saintillan_instabilities_2008}
Saintillan D and Shelley M~J 2008 {\em Phys. Fluids\/} {\bf 20} 123304

\bibitem{goldfriend_hydrodynamic_2016}
Goldfriend T, Diamant H and Witten T~A 2016 {\em Phys. Rev. E\/} {\bf 93}
  042609

\bibitem{jeffrey_calculation_1984}
Jeffrey D~J and Onishi Y 1984 {\em J. Fluid Mech.\/} {\bf 139} 261--290

\end{thebibliography}
		\bibliographystyle{iopart-num}
 
\clearpage
		\appendix
		\section{Traceless symmetric tensors}
		\label{TST}
		Traceless symmetric tensors are commonly used in moment expansions \cite{ahmadi_hydrodynamics_2006}. Starting from the second order nematic tensor
		\begin{equation}
		\label{rank2}
		Q^{\alpha \beta}_{ij} = \int \left(u^{\alpha}_i u^{\beta}_j- \frac{1}{3} \delta_{ij}\delta^{\alpha \beta}\right)  \mathcal{P};
		\end{equation}
		at third order
		\begin{equation}
		\label{rank3}
		Q^\alpha_{ijk} = \int [u^\alpha_{i} u^\alpha_{j} u^\alpha_{k} - \frac{1}{5}(\delta_{ij}u^\alpha_{k} + \delta_{ik}u^\alpha_{j} + \delta_{jk}u^\alpha_{i})]\mathcal{P};\qquad
		T^{\alpha \beta \gamma}_{(ij)k} = \int[u^\alpha_{i} u^\beta_{j}u^\gamma_{k} - \frac{1}{3}u^\delta_{k} \delta_{ij}\delta^{\alpha \beta}]\mathcal{P};
		\end{equation}
		at order four
		\begin{equation}
		\label{rank4}
		Q^\alpha_{ijkl} = \int [u^\alpha_{i} u^\alpha_{j} u^\alpha_{k} u^\alpha_{l} - \frac{1}{15}(\delta_{ij}\delta_{kl} + \delta_{ik}\delta_{jl} + \delta_{il}\delta_{jk})]\mathcal{P};\qquad
		T^{\alpha \beta \gamma \delta}_{(ij)(kl)} = \int [u^\alpha_{i} u^\beta_{j} u^\gamma_{k} u^\delta_{l} - \frac{1}{9}\delta_{ij}\delta_{kl}\delta^{\alpha \beta}\delta^{\gamma \delta}]\mathcal{P};
		\end{equation}
		and finally at order five
		\begin{eqnarray}
		\label{rank5}
		Q^\alpha_{ijklm} & = & \int \bigg\{u^\alpha_{i} u^\alpha_{j} u^\alpha_{k} u^\alpha_{l} u^\alpha_{m} - \frac{1}{21}\big[u^\alpha_m(\delta_{ij}\delta_{kl} + \delta_{ik}\delta_{jl} + \delta_{il}\delta_{jk}) + u^\alpha_l(\delta_{ij}\delta_{km} + \delta_{ik}\delta_{jm} + \delta_{im}\delta_{jk})\nonumber \\
		&& +u^\alpha_k(\delta_{ij}\delta_{lm} + \delta_{il}\delta_{jm} + \delta_{im}\delta_{jl}) + u^\alpha_j(\delta_{ik}\delta_{lm} + \delta_{il}\delta_{km} + \delta_{im}\delta_{kl})+u^\alpha_i(\delta_{jk}\delta_{lm} + \delta_{jl}\delta_{km} + \delta_{jm}\delta_{kl})\big]\bigg\}\mathcal{P} \nonumber \\
		T^{\alpha \beta \gamma \delta \epsilon}_{(ijkl)m} & = & \int [u^\alpha_{i} u^\beta_{j} u^\gamma_{k} u^\delta_{l} u^\epsilon_{m} - \frac{1}{15}u^\epsilon_m(\delta_{ij}\delta_{kl} + \delta_{ik}\delta_{jl} + \delta_{il}\delta_{jk})\delta^{\alpha \beta \gamma \delta}]\mathcal{P}; \nonumber \\
		T^{\alpha \beta \gamma \delta \epsilon}_{(ijk)(lm)} & = & \int [u^\alpha_{i} u^\beta_{j} u^\gamma_{k} u^\delta_{l} u^\epsilon_{m} - \frac{1}{15}(u^\gamma_{k}\delta_{ij} + u^\beta_{j}\delta_{ik} + u^\alpha_{i}\delta_{ik})\delta_{lm}\delta^{\alpha \beta \gamma}\delta^{\delta \epsilon}]\mathcal{P}; \nonumber \\
		T^{\alpha \beta \gamma \delta \epsilon}_{(ij)(kl)m} & = & \int [u^\alpha_{i} u^\beta_{j} u^\gamma_{k} u^\delta_{l} u^\epsilon_{m} - \frac{1}{9}u^\epsilon_{m}\delta_{ij}\delta_{kl}\delta^{\alpha \beta}\delta^{\gamma \delta}]\mathcal{P},\nonumber\\
		\end{eqnarray}
		the highest order encountered in our moment expansion. Parentheses denotes symmetry in the isolated subscripts and the tensors are traceless on contraction of those indices. The sum over Greek letters runs over $0,1$ and $2$, where we have used the notation $\uu^0=\nn$.
\end{document}